\documentclass[twocolumn]{aastex631} 

\usepackage{gensymb}
\usepackage[normalem]{ulem}
\usepackage{hyperref}

\usepackage{graphicx} 

\received{--}
\revised{--}
\accepted{--}
\submitjournal{ApJ}

\newcommand{\WHL}{WHL0137$-$08 }
\newcommand{\SPT}{SPT0615$-$57 }
\newcommand{\MACS}{MACS0308+26 }
\newcommand{\HST}{{\it HST }}

\newcommand{\zphot}{\ensuremath{z_{\text{phot}}}}

\shorttitle{Lensed Arcs}
\shortauthors{Welch et al.}

\begin{document}

\title{RELICS: Small-scale Star Formation in Lensed Galaxies at $z = 6 - 10$}

\correspondingauthor{Brian Welch}
\email{bwelch7@jhu.edu}

\author[0000-0003-1815-0114]{Brian Welch}
\affiliation{Center for Astrophysical Sciences, Department of Physics and Astronomy, The Johns Hopkins University, 
3400 N Charles St. 
Baltimore, MD 21218, USA}

\author[0000-0001-7410-7669]{Dan Coe}
\affiliation{Space Telescope Science Institute (STScI), 
3700 San Martin Drive, 
Baltimore, MD 21218, USA}
\affiliation{Association of Universities for Research in Astronomy (AURA) for the European Space Agency (ESA), STScI, Baltimore, MD, USA}

\author[0000-0002-0350-4488]{Adi Zitrin}
\affiliation{Physics Department,
Ben-Gurion University of the Negev, P.O. Box 653,
Be'er-Sheva 84105, Israel}

\author[0000-0001-9065-3926]{Jose M. Diego}
\affiliation{Instituto de F\'isica de Cantabria (CSIC-UC). Avda. Los Castros s/n. 39005 Santander, Spain}

\author[0000-0001-8156-6281]{Rogier Windhorst}
\affiliation{School of Earth and Space Exploration, Arizona State University, Tempe, AZ 85287, USA}

\author{Nir Mandelker}
\affiliation{Centre for Astrophysics and Planetary Science, Racah Institute of Physics, The Hebrew University, Jerusalem, 91904, Israel}

\author[0000-0002-5057-135X]{Eros Vanzella}
\affiliation{INAF – OAS, Osservatorio di Astrofisica e Scienza dello Spazio di Bologna, via Gobetti 93/3, I-40129 Bologna, Italy}

\author[0000-0002-5269-6527]{Swara Ravindranath}
\affiliation{Space Telescope Science Institute (STScI), 
3700 San Martin Drive, 
Baltimore, MD 21218, USA}

\author[0000-0003-1096-2636]{Erik Zackrisson}
\affiliation{Observational Astrophysics, Department of Physics and Astronomy, Uppsala University, Box 516, SE-751 20 Uppsala, Sweden}

\author[0000-0001-5097-6755]{Michael Florian}
\affiliation{Department of Astronomy, Steward Observatory, University of Arizona, 933 North Cherry Avenue, Tucson, AZ 85721, USA}

\author[0000-0002-7908-9284]{Larry Bradley}
\affiliation{Space Telescope Science Institute (STScI), 
3700 San Martin Drive, 
Baltimore, MD 21218, USA}

\author[0000-0002-7559-0864]{Keren Sharon}
\affiliation{Department of Astronomy, University of Michigan, 1085 S. University Ave, Ann Arbor, MI 48109, USA}

\author[0000-0001-5984-0395]{Maru{\v s}a Brada{\v c}}
\affiliation{University of Ljubljana, Department of Mathematics and Physics, Jadranska ulica 19, SI-1000 Ljubljana, Slovenia}
\affiliation{Department of Physics and Astronomy, University of California Davis, 1 Shields Avenue, Davis, CA 95616, USA}

\author[0000-0002-7627-6551]{Jane Rigby}
\affiliation{Observational Cosmology Lab, NASA Goddard Space Flight Center, Greenbelt, MD 20771, USA}

\author[0000-0003-1625-8009]{Brenda Frye}
\affiliation{Department of Astronomy, Steward Observatory, University of Arizona, 933 North Cherry Avenue, Tucson, AZ 85721, USA}

\author{Seiji Fujimoto}
\affiliation{Cosmic Dawn Center (DAWN), Copenhagen, Denmark}
\affiliation{Niels Bohr Institute, University of Copenhagen, Jagtvej 128, Copenhagen, Denmark}

\begin{abstract}

Detailed observations of star forming galaxies at high redshift are critical to understand the formation and evolution of the earliest galaxies. 
Gravitational lensing provides an important boost, allowing observations at physical scales unreachable in unlensed galaxies. 
We present three lensed galaxies from the RELICS survey at $\zphot = 6 - 10$, including the most highly magnified galaxy at $\zphot \sim 6$ (WHL0137-zD1, dubbed the Sunrise Arc), the brightest known lensed galaxy at $\zphot \sim 6$ (MACS0308-zD1), and the only spatially resolved galaxy currently known at $\zphot \sim 10$ (SPT0615-JD). 
The Sunrise Arc contains seven star-forming clumps with delensed radii as small as 3 pc, the smallest spatial scales yet observed in a $z>6$ galaxy, while SPT0615-JD contains features measuring a few tens of parsecs. 
MACS0308-zD1 contains a $r\sim 30$ pc clump with a star formation rate (SFR) of $\sim 3 M_{\odot} \textrm{ yr}^{-1}$, giving it a SFR surface density of $\Sigma_{SFR} \sim 10^3 M_{\odot}\textrm{ yr}^{-1}\textrm{ kpc}^{-2}$.
These galaxies provide a unique window into small scale star formation during the Epoch of Reionization.
They will be excellent targets for future observations with JWST, including one approved program targeting the Sunrise Arc.

\end{abstract}


\section{Introduction}

Deep field observations with the Hubble Space Telescope (HST) have revealed that galaxies at high redshift tend to be smaller \citep{Shibuya15,Shibuya19,Mowla19,Neufeld21} and exhibit clumpier structures \citep[][]{Shibuya16} than local galaxies. 
In field galaxies, these clump structures were found to have typical radii of $\sim 1$ kpc \citep[e.g.,][]{Elmegreen05,Elmegreen07,Elmegreen09,Guo11,Genzel11,ForsterSchreiber11,Guo15,Guo18}.
These observations were largely limited by the resolution of HST, which cannot observe smaller scales at high redshift without assistance.
The magnifying effect of gravitational lensing has opened a new window into small scale star formation in distant galaxies.
Using HST and strong lensing, many studies have been able to push to scales of $\sim 100$ pc across a broad range of redshifts \citep[e.g.,][]{Jones10,Livermore12,Wuyts14,Livermore15}.
In certain cases, smaller structures can be observed when galaxies reach particularly high magnifications, leading to detections of clumps measuring tens of pc in radius \citep{Vanzella17a,Vanzella15b,Zitrin11_z5clumps,Johnson17L,Zick2020}.
More recently, observations of the highly magnified Sunburst Arc \citep{RiveraThorsen17_sunburst,RivThor19_sunburst2} have revealed star forming clumps as small as $\sim 3$ pc \citep{Vanzella22_sunburst}.
Other observations of lensing cluster MACS J0416 have revealed clumps as small as $\sim 4$ pc \citep{Mestric22}.
At redshifts above $z \sim 1.5$, these small spatial scale observations are further aided by the decreasing angular diameter distance.

These observations of smaller structures have put pressure on the dominant explanation of clump formation in the high-redshift universe. 
Recent studies have found evidence of bias towards larger clump radii and masses at low spatial resolution \citep{Dessauges-Zavadsky17,Cava18}. 
Furthermore, while lower-resolution simulations tended to favor larger, more massive ($\sim 10^9 M_{\odot}$) clumps \citep{Mandelker14}, more recent higher-resolution simulations have found broader mass ranges, and tend to favor smaller ($\sim 10^7 M_{\odot}$) clumps \citep{Tamburello15,Mandelker17,Oklopcic17}.
Thus, continued study of strongly magnified galaxies at high redshift is critical to our understanding of early galaxy structures.

Recently, the study of highly magnified clumps has been pushed to even higher redshift, with \cite{Vanzella19_13pc} reporting a clump measuring 13 pc at $z\sim 6$.
These highly magnified clumps allow exploration of spatial scales otherwise unreachable with current telescopes. 
In particular, they enable studies of clumps on the same scale as local young massive star clusters \citep[YMCs,][]{Portegies-zwart2010_YMCs}.
Recent studies of YMCs in local galaxies have found peaks in the distribution of radii around 2-3 pc, with some examples reaching 10 pc in radius \citep{Bastian12,Ryon17}.
Local observations of globular clusters have found similar distributions of radii \citep{Puzia14}.
Thus YMCs have been proposed as candidate proto-globular clusters, although this remains uncertain \citep[e.g.,][]{BastianLardo18,Kim18,Terlevich18,Kruijssen15,Kruijssen14}. 
Highly magnified objects in the distant universe present an opportunity to directly study these possible globular cluster progenitors.

In this paper, we present HST observations of three highly magnified lensed galaxies at $6 < z \lesssim 10$. 
These galaxies all exhibit clumpy morphologies, and the high magnifications allow us to study them in detail.

In Section \ref{sec:data} we present the HST data used in this study.
Section \ref{sec:lensmodels} presents the lens models used, while Section \ref{sec:clumpmodel} discusses our measurements of clump radii.
Spectral energy distribution (SED) fitting is presented in Section \ref{sec:sed}.
We present our results in Section \ref{sec:results}, and we contextualize these results in the subsequent Section \ref{sec:discussion}.
Finally, we summarize our conclusions in Section \ref{sec:conclusions}.
We assume a flat cosmology with $\Omega_m = 0.3$, $\Omega_{\Lambda} = 0.7$, and $H_0 = 70 \textrm{ km s}^{-1} \textrm{ Mpc}^{-1}$ throughout.

\section{Data}
\label{sec:data}

\begin{figure*}
    \centering
    \includegraphics[width=0.95\textwidth]{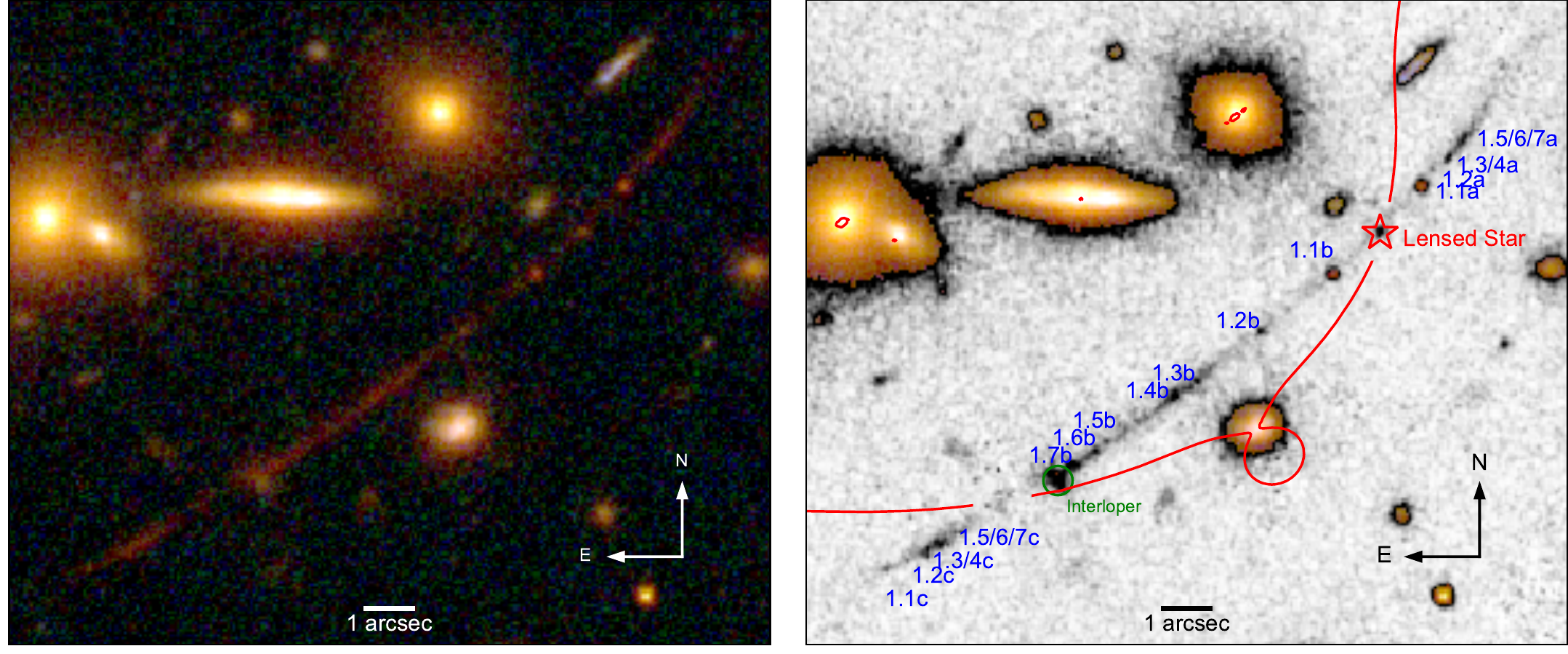}
    \caption{The 15\arcsec\ long Sunrise Arc at $z_{phot} = 6.2 \pm 0.1$ is the longest lensed arc at $z > 6$. 
    The left panel shows a color composite of the arc, with the HST F435W band represented in the blue channel, a combination of F606W and F814W in the green channel, and the full WFC3/IR stack (F105W $+$ F110W $+$ F125W $+$ F140W $+$ F160W) shown in the red channel. 
    The right panel shows a hybrid F110W/color image, with the F110W image shown in pixels with flux below 0.05 e$^{-}$ sec$^{-1} \sim 3.4$ nJy, while pixels above that threshold show the same color image as in the left panel.
    The arc contains seven total clump features labeled in blue in the right panel, with multiple images identified by the letters a/b/c.
    The lensing critical curve is shown in red, broken where it crosses the arc for clarity.
    One lower-redshift interloper appears along the arc, and is circled in green.
    This arc also contains an extremely magnified star, highlighted with a red star, which is discussed in detail in \cite{Welch2022_earendel}.
    Each of the seven lensed clumps have measured radii less than 10 parsecs, providing a detailed look into the substructure of this galaxy. }
    \label{fig:sunrise-arc}
\end{figure*}


The high-redshift galaxies considered here were discovered in the Reionization Lensing Cluster Survey (RELICS; \citealt{Coe19_relics}). 
RELICS observed a total of 41 galaxy clusters with HST, obtaining single-orbit depth in each of the ACS F435W, F606W, and F814W bands along with a total of two orbits split between the WFC3/IR filters F105W, F125W, F140W, and F160W. 
The RELICS clusters were also observed with the Spitzer Space Telescope through the Spitzer-RELICS program (S-RELICS, PI Brada\v{c}).

WHL J013719.8-082841 (henceforth \WHL) was discovered in SDSS imaging as an overdensity of red galaxies by \cite{WHL12}, and its redshift was subsequently confirmed at $z = 0.566$ by \cite{WenHan15}. 
This cluster was ranked as the 31st most massive cluster in the Planck PSZ2 catalog with a Sunyaev-Zeldovich calculated mass of $M_{500} = 9\times 10^{14} M_{\odot}$.  
\cite{Salmon2020} discovered at 15\arcsec\ long arc at $z_{phot} = 6.2$ lensed by this cluster and dubbed it the Sunrise Arc, prompting follow-up observations with HST (GO 15842, P.I. Coe). 
These follow-up images included 5 additional orbits of ACS F814W imaging, and two orbits each of ACS F475W and WFC3/IR F110W imaging, and are presented in detail in \cite{Welch2022_earendel}. 
The Sunrise Arc is shown in Figure \ref{fig:sunrise-arc}, and each of its 7 small clump structures are labeled. 
Additionally, this arc contains an extremely magnified lensed star presented in \cite{Welch2022_earendel}.
\WHL received a total of 7 hours of observations in the IRAC 3.6 $\mu$m band, and 5 hours in the 4.5$\mu$m band.

\begin{figure}
    \centering
    \includegraphics[width=0.45\textwidth]{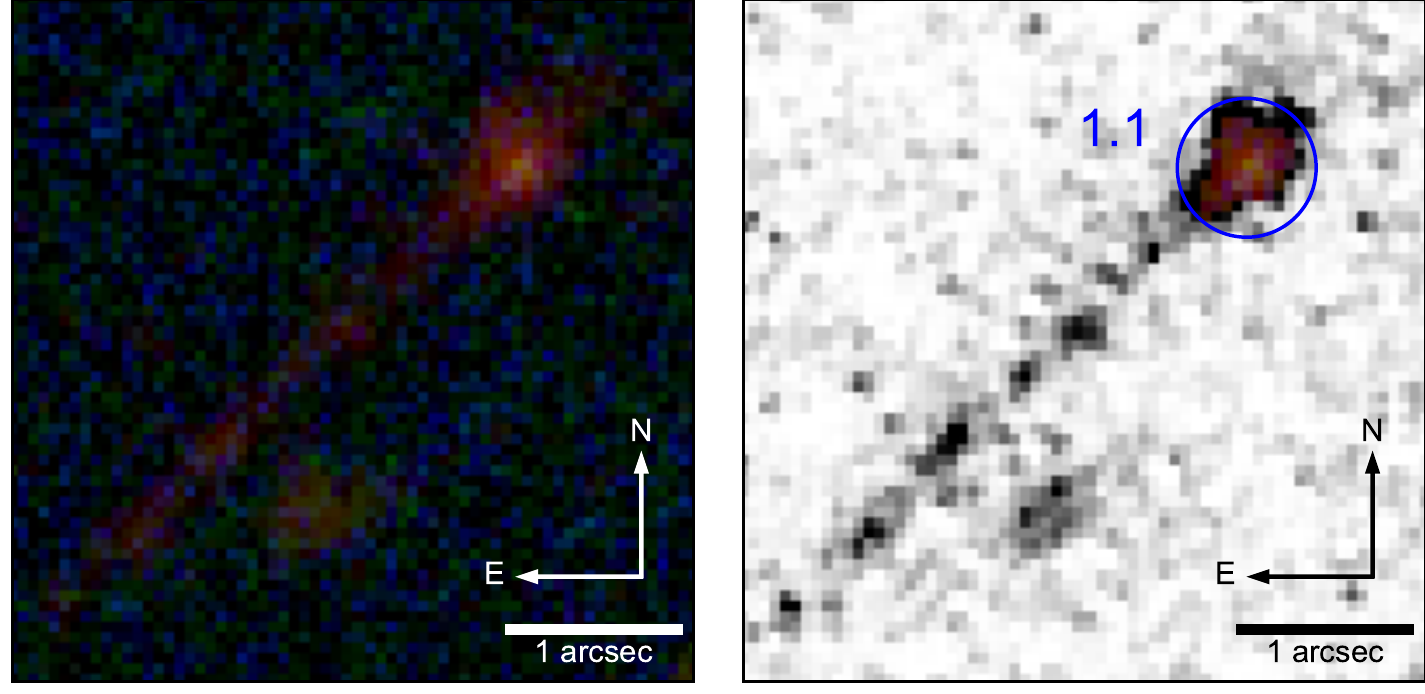}
    \caption{The lensed arc MACS0308-zD1 at $z_{phot} = 6.3 \pm 0.1$ is the brightest known lensed galaxy at $z > 6$. 
    As in Figure \ref{fig:sunrise-arc}, the left panel shows an HST color composite, where the blue channel shows F435W, green is F606W $+$ F814W, and red is the WFC3/IR stack of F105W $+$ F125W $+$ F140W $+$ F160W.
    The right panel shows the hybrid F105W/color image, with a transition threshold of 0.05 e$^-$ sec$^{-1} \sim 7.6$ nJy, similar to Figure \ref{fig:sunrise-arc}.
    The clump structures input into our forward model are labeled in blue in the right panel.
    The brightest clump at the head of the arc (labeled 1.1) has an observed AB magnitude of $\sim 23$. Additional substructures may appear with deeper follow-up imaging.
    The object slightly offset from the tail of the arc is a foreground object, likely either at $z \sim 4$ or a dwarf cluster member galaxy. The bright head of the arc lies about 4.3\arcsec south of the lensing critical curve. }
    \label{fig:macsarc}
\end{figure}

The galaxy cluster MACS J0308+2645 (henceforth \MACS) was presented as part of the Massive Cluster Survey (MACS; \citealt{Ebeling01_MACS}). 
It is at redshift $z = 0.356$ and has an SZ mass of $M_{500} = 10.8\times 10^{14} M_{\odot}$, making it the 12th most massive cluster in the Planck PSZ2 cluster catalog \citep{PSZ2}. 
This cluster lenses an exceptionally bright galaxy dubbed MACS0308-904, which is the brightest lensed arc in the RELICS sample with an AB magnitude of 23.4 in the F160W filter \citep{Salmon2020}. 
This arc is shown in Figure \ref{fig:macsarc}, which and the brightest clump is labeled (1.1).
Additional substructure may be present within the arc, but this is too faint to measure at present.
The cluster \MACS was observed for a total of 5 hours in each Spitzer IRAC band. 

\begin{figure}
    \centering
    \includegraphics[width=0.45\textwidth]{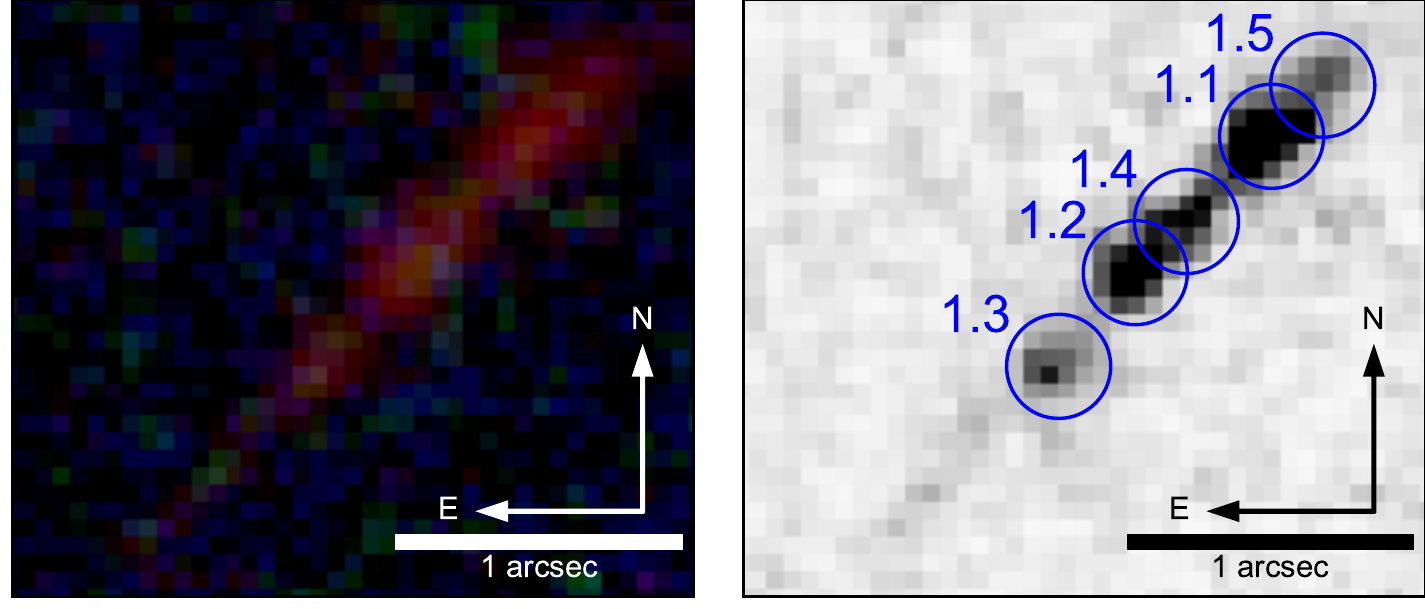}
    \caption{The $\sim 2.5$\arcsec \ long arc SPT0615-JD is the most distant resolved arc observed so far, with multiple photometric redshift estimates putting it solidly at $z\sim 10$. 
    The left panel shows a color composite from HST data, as in previous figures. 
    The right panel shows the F160W band, with each substructure circled and labeled in blue.
    We identify a total of five clump features within this arc, labeled 1.1 through 1.5. 
    Clumps 1.1 and 1.5, as well as clumps 1.2 and 1.4, appear blended in our HST images, due to their proximity and resolution limits. 
    This arc is located about 9\arcsec from the lensing critical curve at $z \sim 10$.}
    \label{fig:spt_arclabel}
\end{figure}

\cite{Salmon18_z10} discovered a $\sim 2.5$\arcsec\ lensed arc of a galaxy with a photometric redshift of $z_{phot} = 9.9^{+0.8}_{-0.6}$ magnified by the galaxy cluster SPT-CL J0615--5746 (hearafter \SPT). 
The lensed arc is dubbed SPT0615-JD1, and it consists of a total of 5 clumps labeled in Figure \ref{fig:spt_arclabel}.
The cluster was discovered independently by the South Pole Telescope survey (SPT; \citealt{Williamson11_spt0615}) and the \cite{Planck11_spt0615}, and it was determined to have a high mass ($M_{500} = 7\times 10^{14} M_{\odot}$) and high redshift ($z = 0.972$).
SPT0615--57 was followed up with additional HST imaging (GO 15920, P.I. Salmon), which obtained an additional one orbit each in F105W and F125W, and two orbits each in F140W and F160W. 
Because of the promising $z\sim 10$ arc in this field, this cluster received additional Spitzer observations, for a total of 30 hours of exposure time in each of the IRAC bands ($3.6 \mu m$ and $4.5\mu m$). 

\begin{table*}[]
    \centering
    \begin{tabular}{c c c c c c}
         Filter & Sunrise Arc & Sunrise Arc 1.1 & MACS0308-zD1 & MACS0308-zD1 1.1 & SPT0615-JD1 \\
         & Flux (nJy) & Flux (nJy) & Flux (nJy) & Flux (nJy)& Flux (nJy) \\
         \hline 
         F435W & $-69 \pm 56$   & $-2 \pm 23$ & $184 \pm 64$ & $88 \pm 50$   & $-27 \pm 20$ \\
         F475W & $16 \pm 27$    & $-10 \pm 11$ &  &  &  \\
         F606W & $-51 \pm 33$   & $6 \pm 14$ & $-19 \pm 34$ & $-44 \pm 27$  & $32 \pm 12$ \\
         F814W & $312 \pm 21$   & $45 \pm 9$ & $426 \pm 50$ & $325 \pm 39$  & $-13 \pm 10$ \\
         F105W & $1321 \pm 74$  & $290 \pm 30$ & $2250 \pm 61$ & $1740 \pm 49$ & $21 \pm 16$ \\
         F110W & $1187 \pm 21$  & $225 \pm 8$ &  &  &  \\
         F125W & $1351 \pm 137$ & $260 \pm 50$ & $2483 \pm 104$ & $1901 \pm 81$ & $70 \pm 15$ \\
         F140W & $1197 \pm 109$ & $230 \pm 40$ & $2116 \pm 83$ & $1625 \pm 65$ & $233 \pm 10$ \\
         F160W & $1088 \pm 74$  & $215 \pm 30$ & $2102 \pm 56$ & $1646 \pm 45$ & $352 \pm 13$ \\
         IRAC Ch.1 &  &  &  &  & $29 \pm 9$ \\
         IRAC Ch.2 &  &  &  &  & $50 \pm 15$ \\
    \end{tabular}
    \caption{Observed photometry from HST and Spitzer for each arc and the bright clump of MACS0308-zD1 used for SED fitting. }
    \label{tab:photometry}
\end{table*}

All \HST images were processed and drizzled to 0.06\arcsec pixels as described in \cite{Coe19_relics}. HST photometry for all sources was measured using Source Extractor v2.19.5 \citep{sextractor} following the method detailed in \cite{Coe19_relics}. 
Photometry for each full arc, as well as for the brightest individual clumps of both the Sunrise Arc and MACS0308-zD1, are presented in Table \ref{tab:photometry}.
Disentangling clump fluxes from the background arc can be tricky. 
In this instance, the two brightest clumps are much brighter than the local background arc, making detailed separation of clump from arc less influential. 
We therefore assume that the clump photometry produced from our fiducial Source Extractor parameters is dominated by the clump itself, with negligible contribution from the background arc. 
This assumption would not hold for other clumps, however we only utilize the multiband photometry to fit SEDs to the two brightest clumps. 
A more careful extraction of clump fluxes is therefore not necessary at this time.

Initial redshifts were measured for all RELICS objects using BPZ \citep{Benitez00bpz, Coe06bpz}. 
This technique yields a redshift of $\zphot = 6.3^{+0.2}_{-0.1}$ for MACS0408-zD1, and we adopt a fiducial redshift of $z  6.3$ for all relevant calculations here.
Further SED modeling was done for SPT0615-JD1 in \cite{Salmon18_z10} and again in \cite{Strait20} to further explore its photometric redshift, along with other physical properties. 
Each method yields a $z\sim 10$ solution ($z_{phot} = 9.9^{+0.8}_{-0.6}$ for \citealt{Salmon18_z10}, and $z_{phot} = 10.2^{+1.1}_{-0.5}$ for \citealt{Strait20}). 
We assume a fiducial redshift of $z = 10.0$ for relevant calculations in this analysis.
Spitzer flux measurements are detailed in \cite{Strait20, Strait21}.
While \cite{Strait21} performed many SED fits to high-redshift RELICS galaxies, that paper did not fit the Sunrise Arc or MACS0308-zD1 due to poor constraints on IR fluxes from Spitzer data. 
Additional SED fitting was done for the Sunrise Arc in \cite{Welch2022_earendel}, finding a photometric redshift of $\zphot = 6.2 \pm 0.1$. 
We adopt a fiducial redshift of $z = 6.2$ for relevant calculations for this arc.

\section{Lens Models}
\label{sec:lensmodels}

Proper interpretation of gravitationally lensed galaxies relies on accurate models of the lensing clusters. 
For our present analysis, we utilize published lens models for each lensing cluster, made using the Lenstool \citep{JulloLenstool07,JulloLenstool09} and Light-Traces-Mass (LTM; \citealt{Broadhurst05,Zitrin09,Zitrin15}) lens modeling softwares.

\subsection{WHL0137}
\label{whlmodel}
WHL0137 has a total of five published lens models made using four different lens modeling tools: Lenstool, LTM, Glafic \citep{Oguri2010} and WSLAP+ \citep{Diego05wslap,Diego07wslap2}, the details of which can be found in \cite{Welch2022_earendel}.
For the present analysis, we focus on the LTM model which produces the most accurate reproduction of the full length of the Sunrise Arc. 
While all models produce a reasonably faithful reproduction of the arc, only LTM simultaneously matches the positions and relative brightnesses of all components.
Notably, the Lenstool and Glafic models produce much higher magnifications for clump 1.1b than for clump 1.1a, inconsistent with the observed relative fluxes which are consistent within $1\sigma$ uncertainties in each WFC3/IR band. 
The lens models for WHL0137 are constrained by two photometrically identified multiple image systems: the Sunrise arc at $z \sim 6.2$ as well as a triply-imaged $z \sim 3$ galaxy. 

The LTM model predicts magnifications of $\mu \sim 60 - 250$ for individual clumps along the arc. 
The rapidly changing magnification in the vicinity of lensing critical curves is the cause of this wide range of magnification predictions, and makes setting a fiducial value of magnification for the full arc difficult. 
Where necessary, we adopt a total magnification of $\mu = 155 \pm 13$ for the full arc (inclusive of all three multiple images) and a magnification of $\mu = 130 \pm 70$ for the most highly stretched central image.
The total magnification is calculated using a ratio of observed flux to forward model flux, described in Section \ref{sec:clumpmodel}.

\subsection{MACS0308}

MACS0308 was modeled by \cite{Acebron18_macs0308} using the LTM software. 
The lens model is constrained by three multiple image systems, each photometrically measured at $z \sim 2-3$. 
The bright lensed arc MACS0308-zD1 is not included as a constraint in the lens modeling.

This lens model predicts a lensing magnification of $\mu \sim 20$ for the bright $z\sim 6$ arc. 
We adopt $\mu = 20$ as our fiducial magnification for analysis of this arc, however given its extended morphology that magnification will change across the length of the arc. 
Particularly, magnification decreases to the southeast of the brightest clump, meaning the fainter tail of the arc is at lower magnification than the bright clump. 
While the brightest clump is magnified by a factor 22, the furthest tail of the arc is only magnified by a factor 15.
Its potential counterimage has a model predicted magnification of $\mu \sim 2$.

\subsection{SPT0615}

The lens model for SPT0615 was presented in \cite{Paterno-Mahler18_spt0615} and utilizes the Lenstool modeling software. 
This model is constrained by three multiple image families, two of which include spectroscopic redshifts. 
This lens model predicts a magnification for SPT0615-JD1 of $\mu \sim 4-7$, varying along the length of the arc. 
We adopt a conservative fiducial magnification for the full arc of $\mu = 5$ where necessary. 
The lens model of \cite{Paterno-Mahler18_spt0615} predicts that two additional multiple images of SPT0615-JD1 should be visible.
These multiple images have yet to be identified in the \HST imaging, as they are likely below the detection limits of current observations.

\section{Clump Modelling}
\label{sec:clumpmodel}

Gravitational lensing can provide a significant boost in resolution, allowing us to probe much smaller physical scales than would be possible in field galaxies. In order to derive the greatest benefit from this increase in resolution, we use a forward modeling technique similar to that described in \cite{Johnson17model}. 

In this section, we describe the forward modeling code, as well as the star formation rate estimate we make using its outputs. We then detail the modeling decisions made for each arc individually.

\begin{figure*}
    \centering
    \includegraphics[width=0.9\textwidth]{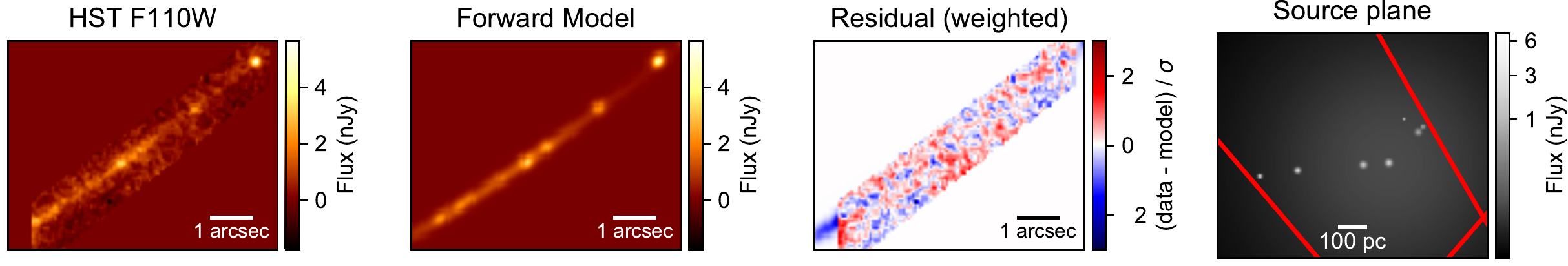}\\
    \includegraphics[width=0.9\textwidth]{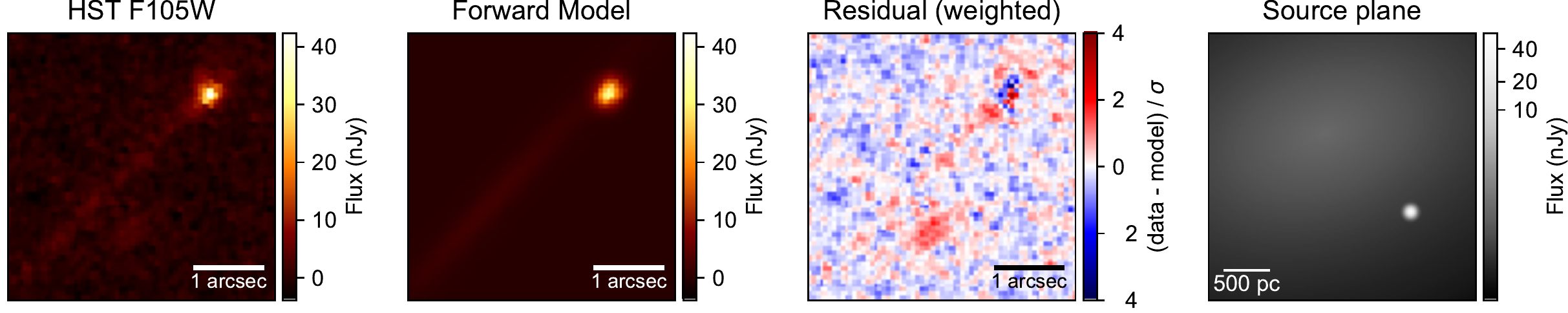}\\
    \includegraphics[width=0.9\textwidth]{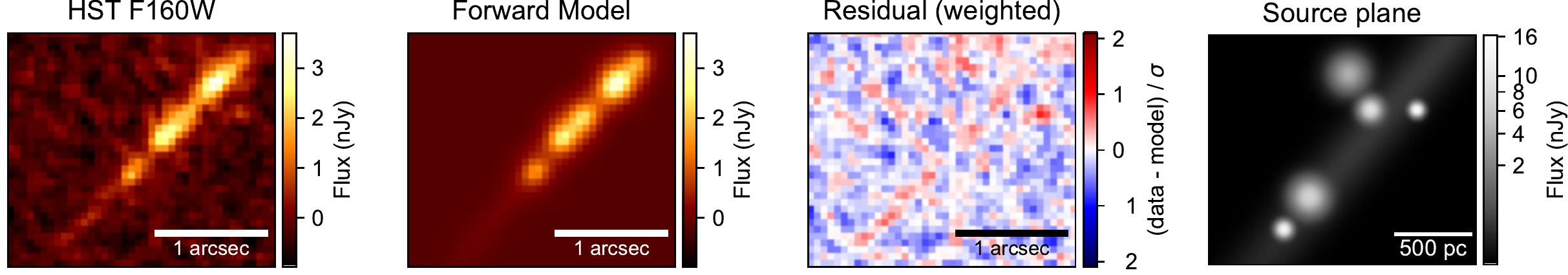}
    \caption{ Forward model fitting allows detailed study of the substructure of lensed galaxies. Here we show HST images of our lensed arcs (left) along with forward model image plane reconstructions of each arc (left middle). The weighted residuals, calculated by subtracting the model from the data and dividing by pixel-level uncertainty, are shown in the middle right column. The residuals are consistent with noise in each fit, indicating that our forward model is successfully capturing the full arc in each case. The source plane models for each galaxy are shown in the right-hand column. Note the image stretch causes the wings of the brighter clumps to be more visible, making light from these clumps visible out to 2-3 times the quoted radius (2-3$\sigma$ in the Gaussian profile). In the top row depicting the Sunrise Arc, the caustic curves are shown in red in the source plane reconstruction. }
    \label{fig:fwdmodel_resid}
\end{figure*}

\subsection{Forward Model}
Our forward modeling technique provides a measurement of the intrinsic size, morphology, and brightness of highly magnified galaxy substructures.

The forward model begins with the creation of a source plane model. 
This model includes a number of predefined clump structures, along with a larger background structure to represent the diffuse light of the host galaxy. 
Each structure is modeled according to a parametric light profile, and is positioned in the source plane based on the delensed centroid in the image plane.
Currently, the code can include 2-dimensional Gaussian profiles, and Sersic profiles.
For this analysis, we primarily model structures with a two-dimensional Gaussian profile. 
Many of the substructures examined here are unresolved or barely resolved, so the simpler Gaussian profile proves sufficient for their characterization. 
For larger resolved structures, Sersic profiles can be used.
We attempted to fit these structures with Sersic profiles, however we generally found that the additional model complexity did not improve the fitting result. 

Once an initial source plane model has been defined, we create an image plane reconstruction using the lens model deflection maps. This image plane model is then convolved with an estimated instrument point spread function (PSF) to match the actual HST data. We estimate the PSF using stars observed within our observations. We create a PSF convolution kernel by averaging several stars to account for variations in subpixel position variations. We intentionally avoid particularly bright stars when generating the PSF kernel, as the diffraction spikes around bright stars would overwhelm the instrument PSF and introduce unnecessary noise in our model. 

To determine the best model parameters, we perform an initial parameter optimization using a downhill simplex algorithm (using Scipy minimize). This optimization provides an estimate of the parameter values, but no estimate of the variance of each parameter. We therefore next use an MCMC to probe the range of possible solutions and estimate the variance on each model parameter. The MCMC is done using the python package emcee \citep{Foreman-Mackey13}.

Before fitting, we set flat priors on the amplitude, width, and ellipticity of each model component. 
The primary function of the priors is to prevent unphysical solutions. 
For example, in the case of unresolved structures, the code can produce arbitrarily small point sources that still reproduce the unresolved image.
To prevent this behavior, we set a lower limit on the Gaussian width based on the magnification and resolution of the HST images.
This cutoff improves our estimates of the upper limits on the radii of unresolved objects. 

Currently, the forward model fit analyzes only a single-band image.
For our analysis, we utilize the bluest filter with sufficient signal to distinguish clump features.
The appropriate filter choice varies between arcs based on redshift and available image depth.
Utilizing the bluest available filter ensures the smallest PSF possible, and thus the most precise measurement of clump radii.


\subsection{Star Formation Rate Calculation}

The forward model provides a pathway to measure star formation rates for clumps via their rest-frame far-UV luminosity. This is particularly useful for faint clumps for which SED fitting does not produce well-constrained results. 

We first use the output of the forward model to calculate the delensed flux density from each clump by integrating over the clump profile. In the case of a Gaussian profile, the flux can be calculated as 
\begin{equation} 
    F = 2\pi A \sigma_x \sigma_y 
\end{equation}
where $A$ is the Gaussian amplitude, and $\sigma_{x,y}$ are the width in either direction. For symmetric Gaussian profiles, $\sigma_x = \sigma_y$. 
We then divide this integrated flux by a factor $(1+z)$ to correct for bandwidth compression. 
From here, we calculate SFRs for each clump using the far-UV luminosity -- SFR conversion from \cite{MnD14}, namely
\begin{equation}
    \textrm{SFR} = 1.15 \times 10^{-28} L_\nu \text{(FUV)}
\end{equation}
where SFR is in units of $M_{\odot} \textrm{ yr}^{-1}$ and $L_\nu$ is in units of ergs s$^{-1} \textrm{ Hz}^{-1}$.
This relation assumes a \cite{SalpeterIMF} IMF, however \cite{MnD14} calculate a conversion to \cite{ChabrierIMF} or \cite{Kroupa01_imf} IMFs can be made by multiplying by a factor $\sim 2/3$. 

Calculating SFR from UV light can be significantly impacted by any amount of dust absorption. 
For these calculations, we assume no dust absorption. 
SED fitting of each galaxy yields fairly low dust attenuation $A_V < 0.1$ mag, indicating that this assumption is reasonable. 

\subsection{Individual Arc Modeling}

\subsubsection{Sunrise Arc}

The Sunrise Arc consists of three multiple images crossing the lensing critical curve in two places. 
The central image has the highest overall magnification as it runs roughly parallel to the critical curve. 
Thus, we use this central image for our forward modeling.
This arc is best detected in the deep F110W imaging, so we use this band in our forward model fitting.

We identify a total of seven clump structures within this arc segment, labeled 1.1 through 1.7 (see Figure \ref{fig:sunrise-arc}). 
Clumps are identified as separate segments in the Source Extractor segmentation maps. 
Segments of each arc are then visually inspected in the highest signal-to-noise image (F110W in the case of the Sunrise Arc).
Each clump appears unresolved in the WFC3/IR imaging, showing no deviation from a PSF-like point source when examined independent of the background arc. 
Thus our radius constraints are upper limits only. 
The arc contains a measurable diffuse component, which we model as a Gaussian with zero ellipticity. 
Given the high lensing shear in this region, only one dimension is probed in great detail, so the round Gaussian profile is sufficient and reduces the total number of parameters needed for the model. 

The length of the arc, coupled with the crowded field in which it is located, lead us to cut out a rectangular region around the arc for fitting purposes. 
This region removes contamination from nearby cluster galaxies, which can overwhelm the fitting code and produce poor results.
We also choose to cut out the lower-redshift interloper that appears at the southeast end of the central image of the arc (see Fig. \ref{fig:sunrise-arc}). 
This interloper would bias the size and brightness measurements for clump 7, causing it to favor a larger and brighter model.

The resulting best fit for this arc produces a residual that is consistent with background noise (Figure \ref{fig:fwdmodel_resid}).

\subsubsection{MACS0308-zD1}

We identify one clump feature in MACS0308-zD1, which is the bright clump at the head of the arc. 
We note that the tail exhibits hints of additional clump structures, and indeed this section of the arc is split into multiple components in the Source Extractor segmentation maps.
However visual inspection determined that the arc tail is too faint to produce reliable clump detections.
Thus we here model the tail of the arc as a single Gaussian profile with ellipticity less than 0.4.
Additional substructure may be present beyond the detection limits of these observations.

We note that the bright clump at the head of the arc appears to have additional structure beyond the Gaussian we use for fitting, particularly a protrusion off the north edge of the clump. 
We did attempt to model this protrusion, however we found it to be highly degenerate with the main clump, leading to increased uncertainties on both clump parameters without a noticeable decrease in residuals.
There also appears to be a small residual near the center of the clump.
We attempted to model this using a Sersic profile, which could better model a more peaked light distribution, however this residual remained. 
We thus choose to only include one Gaussian component to describe the bright clump. 
It is likely that additional substructure exists in this clump, indicating that it is made up of multiple smaller unresolved objects.

\subsubsection{SPT0615-JD1}

We identify a total of five clump structures within the SPT0615-JD1 arc using the WFC3/IR F160W image. 
The arc includes three clear features, however the brightest two of these three features appear asymmetrically bright in the HST images. 
We thus determine that these asymmetric clumps are likely blended images of multiple smaller structures. 
Previous studies of highly lensed galaxies have found that many small clumps will appear as such asymmetric larger structures when simulated at lower lensing magnification, supporting our determination to model the arc as multiple smaller clumps \citep{Rigby17}.
Including the two additional clumps to describe this arc does introduce some degeneracy between clumps, increasing uncertainties in clump properties.
However, the inclusion of these additional structures leads to noticeable improvement in the residuals, indicating that the added complexity and degeneracy does improve the final fit.

There is a diffuse component to this arc, however it is very faint.
We included this in our model as a Gaussian profile, however it is poorly constrained due to its faintness.

\section{SED Fitting}
\label{sec:sed}

\begin{figure*}
    \centering
    \includegraphics[width=0.32\textwidth]{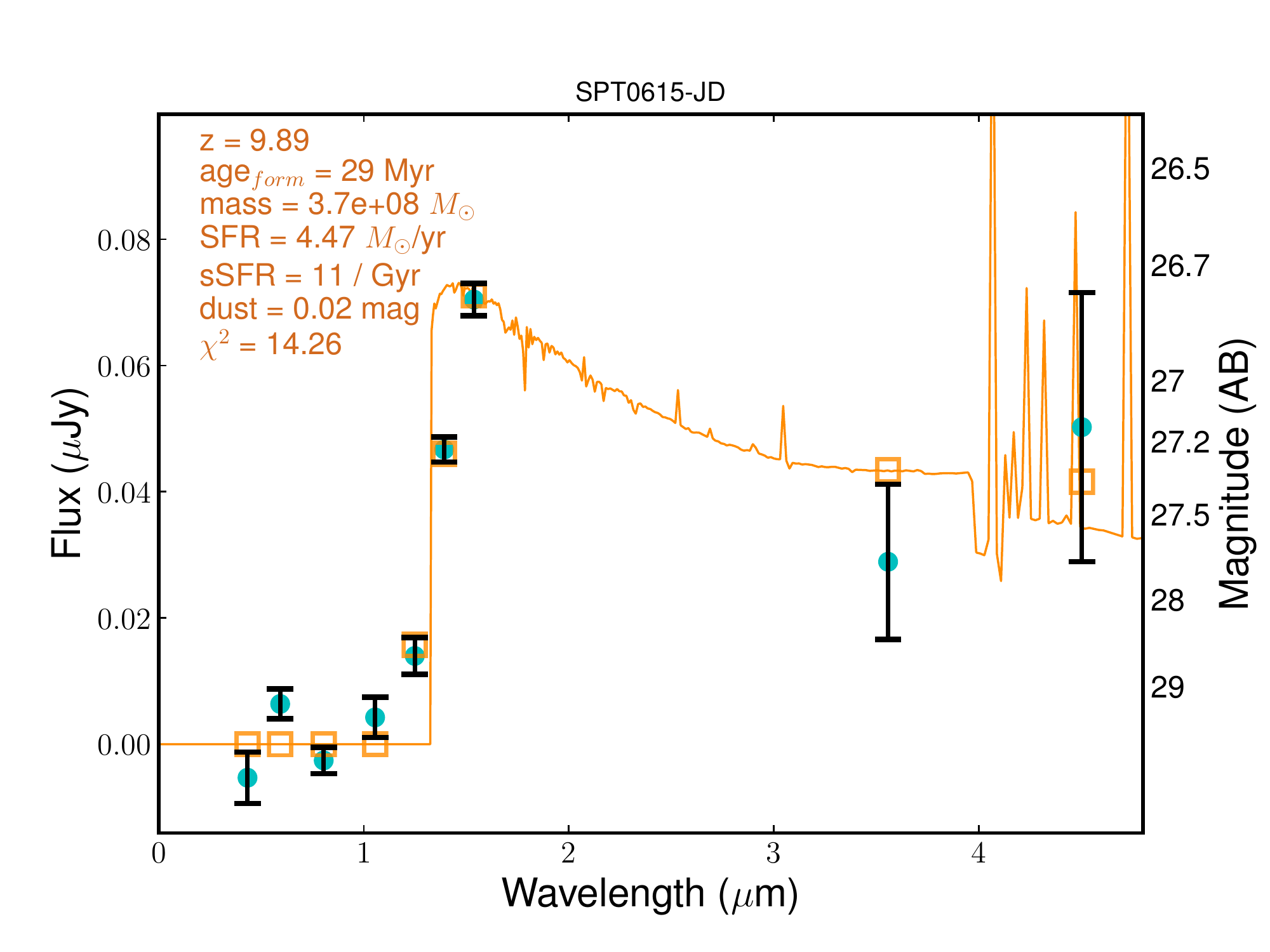}
    \includegraphics[width=0.32\textwidth]{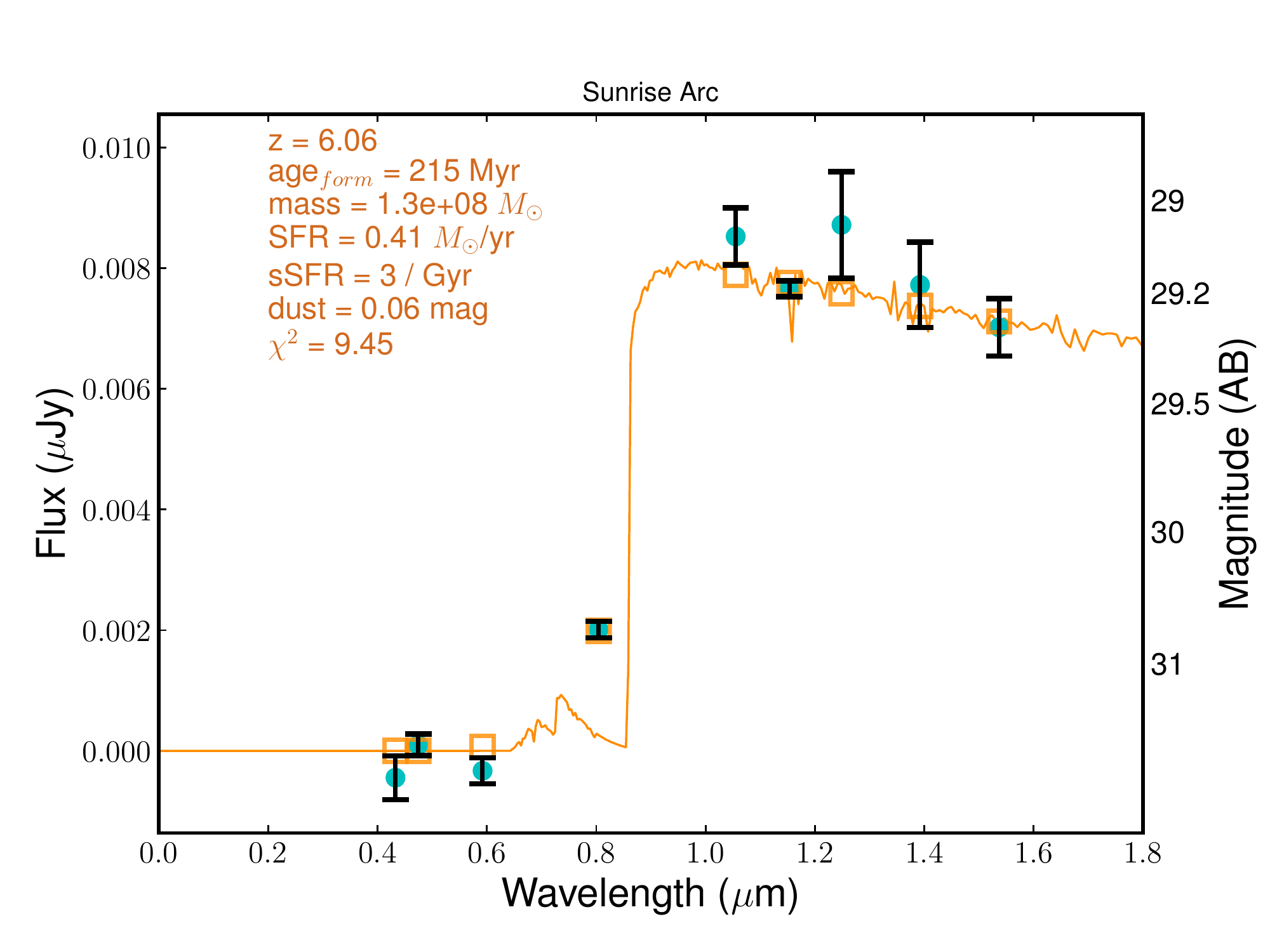} 
    \includegraphics[width=0.32\textwidth]{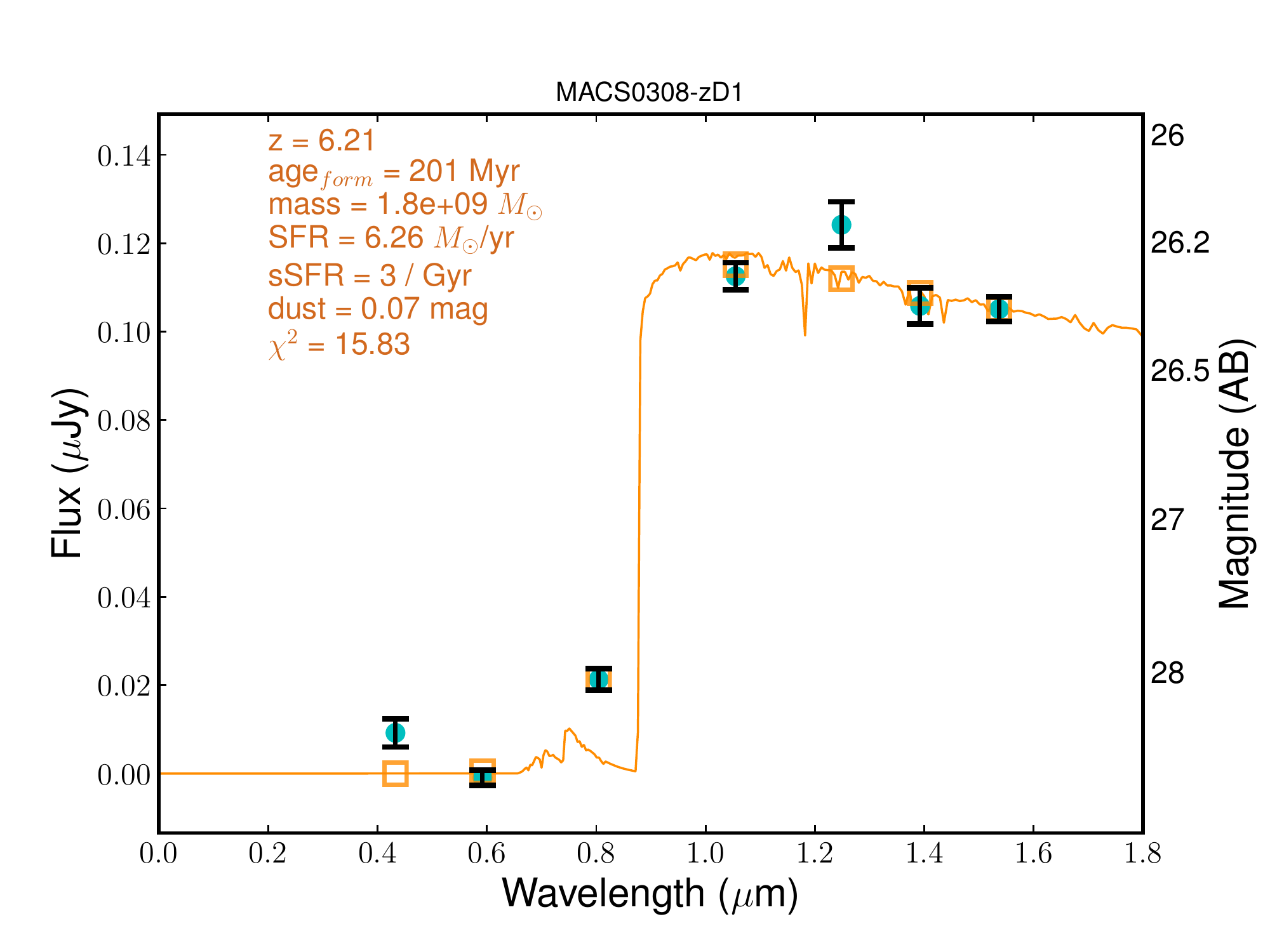}
    \caption{Magnification-corrected SED fits for each galaxy considered are shown. Cyan points with errorbars represent the observed photometry, corrected for magnification. We use our fiducial arc magnifications here, namely $\mu = 155$ for WHL0137-zD1, $\mu = 20$ for MACS0308-zD1, and $\mu = 5$ for SPT0615-JD. The orange line is the best-fit spectrum, and orange boxes are predicted fluxes in each filter based on that best-fit spectrum. With current data, the SFR for each full galaxy can be fairly well constrained. Parameters such as age require additional wavelength coverage to yield strong constraints.}
    \label{fig:sed-fits-full}
\end{figure*}

\begin{figure}
    \centering
    \includegraphics[width=0.45\textwidth]{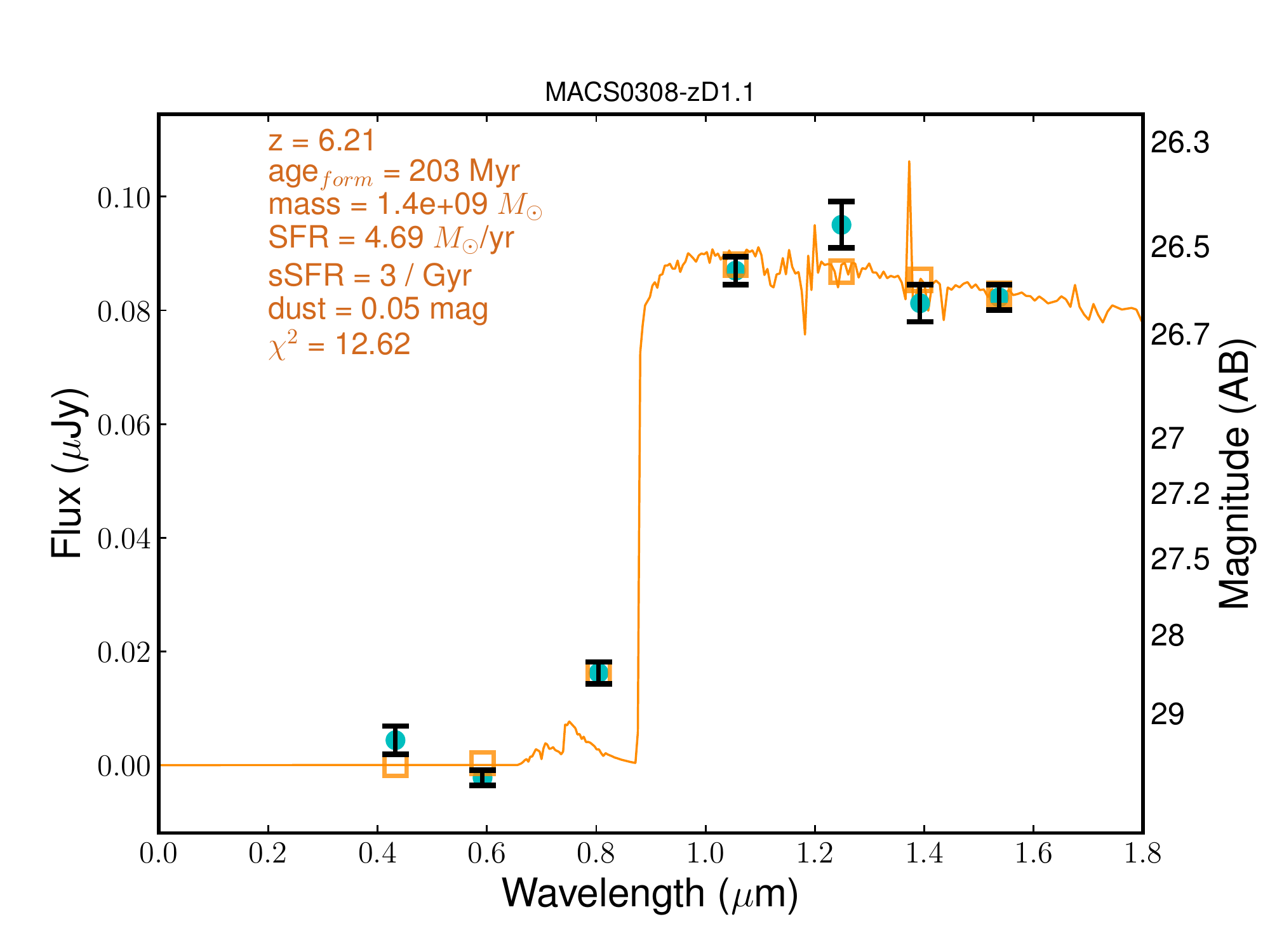}
    \includegraphics[width=0.45\textwidth]{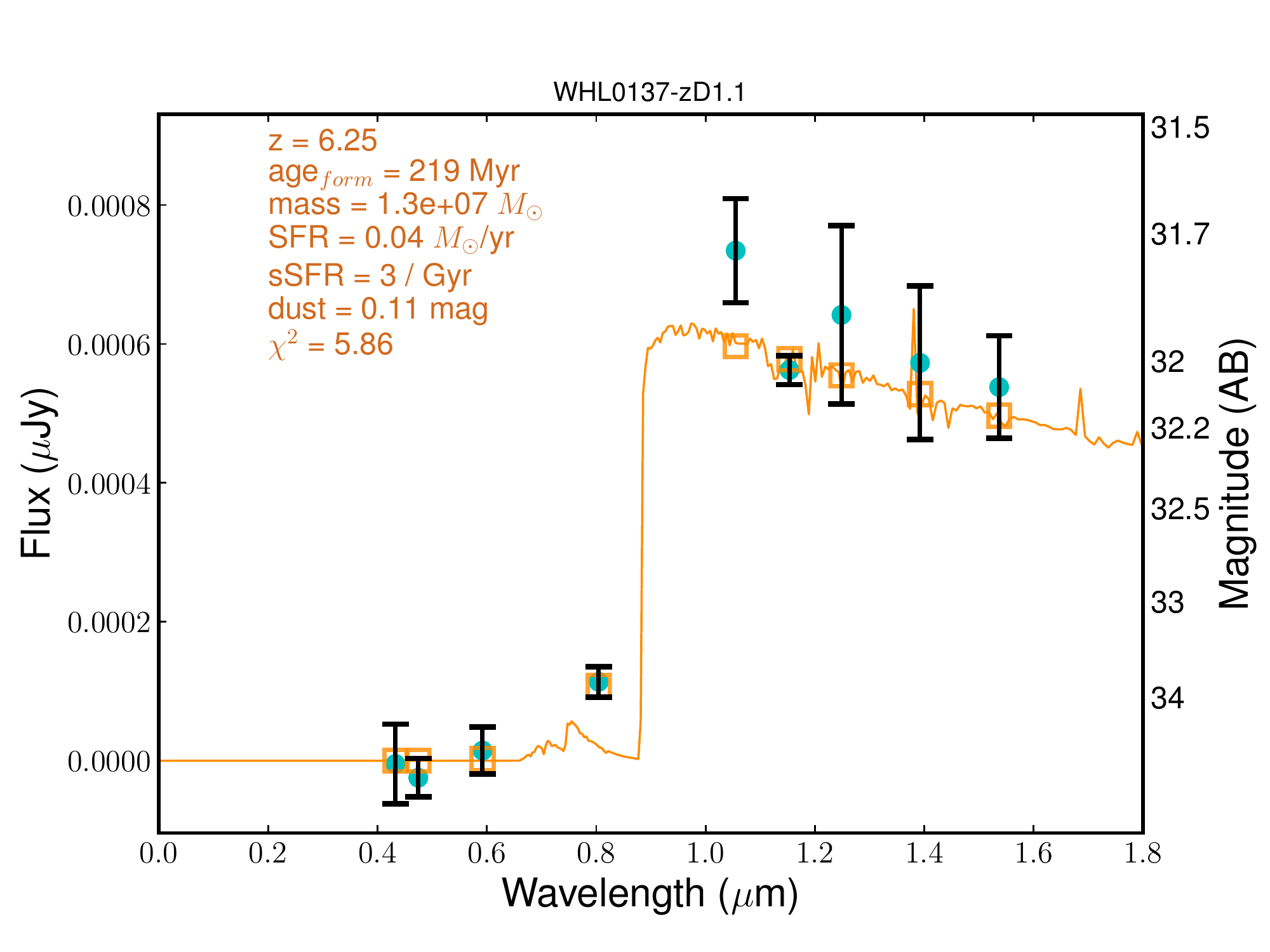}
    \caption{Magnification-corrected SED fits for the brightest clumps within the Sunrise Arc and MACS0308-zD1. Cyan points with errorbars represent the observed photometry, corrected for magnification. For these clumps, we use the magnifications presented in Table \ref{tab:fwd_clumps}, namely $\mu = 215$ for WHL0137-zD1.1, and $\mu = 22$ for MACS0308-zD1.1. The orange line is the best-fit spectrum, and orange boxes are predicted fluxes in each filter based on that best-fit spectrum. Only these two clumps are bright enough to produce reliable SED fits individually. }
    \label{fig:sed-fits-clumps}
\end{figure}

To better understand the properties of the arcs presented here, we perform SED fitting based on available HST and Spitzer photometry.
We use the SED fitting code BAGPIPES \citep{Carnall18_bagpipes} to fit our photometric data, leaving redshift as a free parameter. 
Because each of these objects has a solid photometric redshift estimate in place, we set a redshift range in our fitting of $4 < z < 15$. 
Previous works \citep{Welch2022_earendel,Strait20,Salmon18_z10} have investigated and ruled out lower redshift solutions, justifying our restricted redshift range.
Additionally, attempted fits with fixed redshifts calculated from previous works did not significantly change the results of our SED fits. 

We use the BAGPIPES default stellar population models from \cite{Bruzual_Charlot_03}, coupled with nebular reprocessing implemented by the photoionization code CLOUDY \citep{Ferland17}. 
We model each galaxy with an exponential star formation history of the form $SFR \propto e^{-t/\tau}$, allowing $\tau$ to vary from 100 Myr to 10 Gyr. 
Metallicity is allowed to vary from 0 to $2.5Z_{\odot}$, while stellar mass is left with a wide parameter space of $1 < \log (M_* / M_{\odot}) < 15$. 
Age is varied from 1 Myr to the age of the universe, and ionization parameter is varied over the range $-4 < \log(U) < -2$. 
Dust extinction is implemented using the \cite{Calzetti00} law with $A_V < 3$, and we assume that dust extinction is twice as high around HII regions in the first 10 Myr of their lifetimes. 

Each of these lensed arcs was detected as a series of multiple source segments in Source Extractor, so to perform SED fitting of the full arcs we sum the flux from these segments, as well as summing the uncertainties in quadrature. 
To estimate the intrinsic (delensed) properties of each galaxy, we divide by a fiducial magnification for each arc. 
The fiducial magnifications are discussed in Section \ref{sec:lensmodels}.
The resulting best fit SEDs for each full arc are presented in Figure \ref{fig:sed-fits-full}, along with the photometry for each object. 

We attempt to fit each clump's SED individually, however we find that only the brightest clumps of the Sunrise Arc and MACS0308-zD1 produce reliable fits. 
These best fit SEDs are shown along with the source photometry for each clump in Figure \ref{fig:sed-fits-clumps}.
The remaining clumps of the Sunrise Arc have too low signal to noise to sufficiently constrain the SED, resulting in fractional uncertainties near 100\%.
Meanwhile, the high redshift of SPT0615-JD means that Spitzer IR data is necessary to infer physical parameters from the SED. 
The reduced spatial resolution of the Spitzer images, combined with the faintness of the arc, mean that we cannot reliably extract IR fluxes for individual clumps. 
Meaningful SED fits can therefore not be obtained for clumps within this $z\sim 10$ arc, only the full arc can be reliably fit.


\section{Results}
\label{sec:results}

\begin{figure*}
    \centering
    \includegraphics[width=0.95\textwidth]{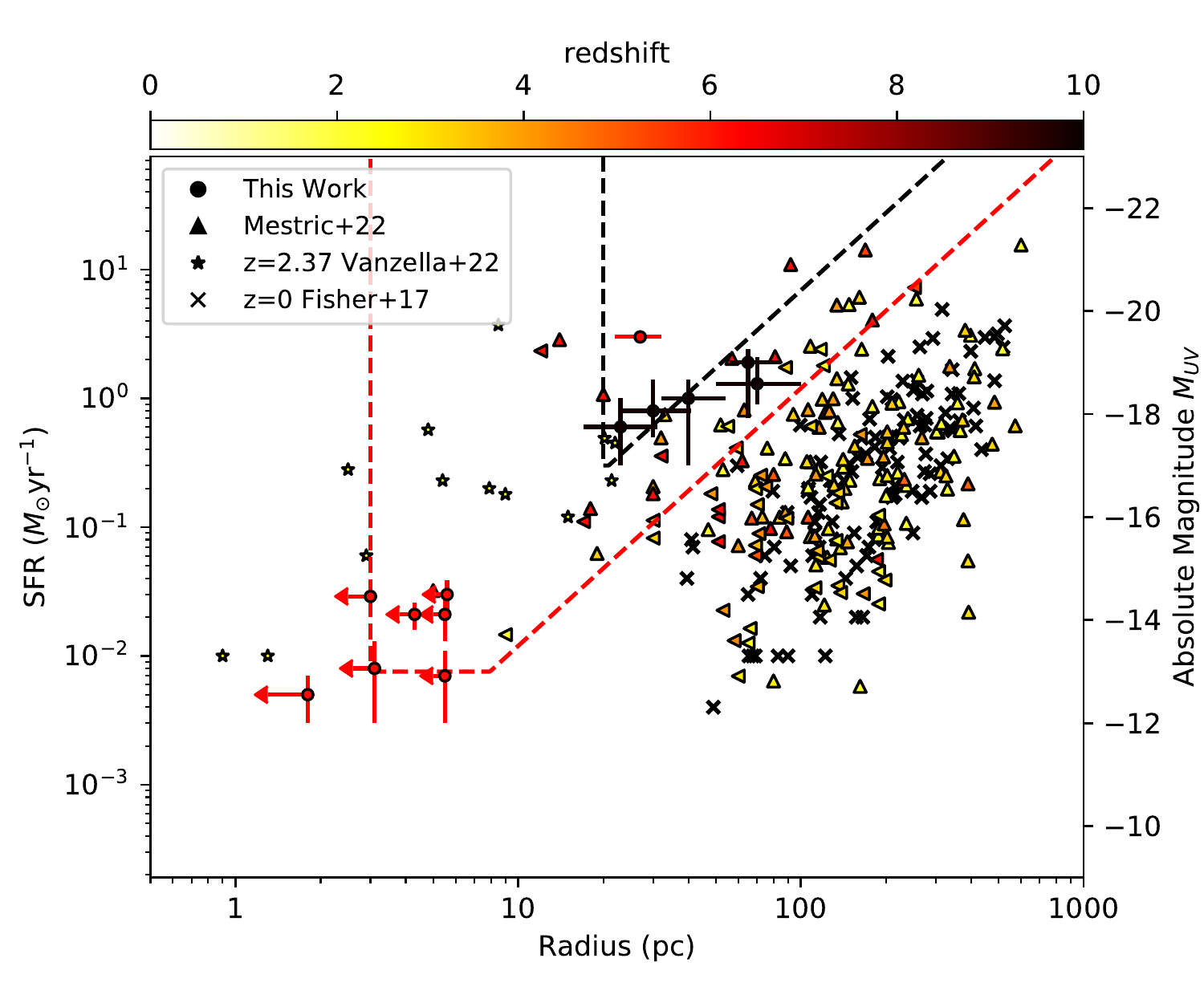}
    \caption{Star formation rates and radii of galaxy substructures from this work (plotted as circles with error bars) as well as several literature samples (triangles: \cite{Mestric22}, stars: \cite{Vanzella22_sunburst}, crosses: \cite{Fisher17}. Note left-facing triangles from the \cite{Mestric22} sample are upper limits on clump radius).
     Colors of each point correspond the the object's redshift, except for the local ($z=0$) sample from \cite{Fisher17}. 
     Additionally, completeness limits are plotted as dashed lines (red for $z = 6$ and black for $z = 10$). The vertical component is the magnification-boosted diffraction limit (note we used the brightest clump of the Sunrise Arc to measure this limit at $z = 6$, and one fainter clump appears at slightly higher magnification in our fiducial lens model). The horizontal line at the bottom of the curve is the point-source sensitivity limit, and the slanted component is the survey surface brightness limit, which has a slope of 5 Mag/dex. These completeness limits indicate that our samples are likely incomplete, and additional substructure may exist beyond the limits of \HST.
     The clumps presented here probe the smallest physical scales yet examined at $z>6$ and $z \sim 10$. Additionally, many of the clumps have particularly high star formation rates given their sizes. These properties are similar to what is seen in local young massive star clusters.}
    \label{fig:sfr_radius}
\end{figure*}

\subsection{Forward Model Radii and SFRs}

Our forward modeling analysis produces direct constraints on clump radii in the source plane, which are presented in Table \ref{tab:fwd_clumps}. 
For the Sunrise Arc, each clump appears unresolved in the image plane (PSF-like), so we can only set upper limits on their sizes. 
The upper limits indicate that these are some of the most compact structures thus far observed at $z \sim 6$, as most have 68\% confidence radius limits less than 5 pc, and 95\% confidence limits less than 10 pc. 
Previously, \cite{Vanzella19_13pc} reported a YMC candidate at $z = 6.143$ with upper limit $r < 13$ pc, meanwhile smaller clumps at scales of a few parsecs have been found in local galaxies \citep{Portegies-zwart2010_YMCs}. 
Radii for clumps in the other arcs are not so compact, they are at scales of a few tens of parsecs each. 
These galaxies are not as highly magnified, meaning they may host smaller star clusters that cannot be resolved with HST imaging. 
In particular, the $z\sim 10$ arc SPT0615-JD1 may include more, smaller clump structures which are blended together to form the larger clumps measured.

Our forward model provides a useful estimate of star formation rates for each clump based on best-fit intrinsic UV brightness. 
The SFRs for each clump are listed in Table \ref{tab:fwd_clumps}, along with their intrinsic UV absolute magnitudes.
The SFR is shown plotted against clump radius in Figure \ref{fig:sfr_radius}, along with a selection of clumps from previous literature.
The literature comparison includes clumps at a range of redshifts and magnifications, including results from galaxies lensed by MACS J0416 \citep{Vanzella21, Mestric22}, clumps within the highly magnified Sunbrust Arc \citep{Vanzella22_sunburst}, and a sample of local ($z = 0$) clumps in nearby galaxies \citep{Fisher17}.
The clumps presented here stand out as some of the smallest clumps known at high redshift, as well as having the highest star formation rates for their size.

From the combination of SFR and clump radius, we can calculate the surface density of star formation 
for each clump as $\Sigma_{\textrm{SFR}} = (0.68 * \textrm{SFR}) / (\pi r^2)$, where the factor 0.68 accounts for the fact that the radius is given by the $\sigma$ of a Gaussian profile, yet the full profile integrated to infinity is used to calculate the SFR.
From this calculation, we see that the $\Sigma_{SFR}$ is highest in the bright clump of MACS0308-zD1.1, followed by the brightest clump of the Sunrise Arc (1.1). 
Each of these show SFR densities greater than $1000\,M_{\odot} \textrm{ yr}^{-1} \textrm{ kpc}^{-2}$, notably higher than other clumps analyzed here, indicating that these are particularly dense star forming systems. 
The remainder of the clumps have SFR densities of order a few hundred $M_{\odot} \textrm{ yr}^{-1} \textrm{ kpc}^{-2}$.
It is worth noting that the values of $\Sigma_{SFR}$ for the Sunrise Arc clumps are best interpreted as lower limits, as the clumps are unresolved in HST imaging. 
If the true radii of these clumps turn out to be smaller than the constraints presented here, they will in turn show higher SFR surface densities.

\subsection{BAGPIPES Results}

We can further assess physical parameters of these galaxies using SED fits. 
We fit the combined flux of each arc using BAGPIPES, and the resulting parameters are presented in Table \ref{tab:pipes_clumps}. 
We attempted to fit each clump individually, however this only worked for MACS0308-zD 1.1 and WHL0137-zD1.1.
The rest of the clumps of the Sunrise Arc proved too faint to yield reliable results from SED fitting. 
Meanwhile, the clumps of SPT0615-JD1 rely heavily on Spitzer photometry to constrain the SED.
The Spitzer resolution is too low to reliably distinguish individual clumps, preventing clump-by-clump SED fitting.


The SED fitting provides estimates of stellar mass, which the forward modeling cannot.
The mass measurements for both WHL0137-zD1.1 and MACS0308-zD1.1 suggest these clumps are particularly dense, with a high stellar mass packed into a small region.
Using the combination of stellar mass and radius, we can calculate a crossing time for each clump following Equation 1 of \cite{Gieles11}.
The crossing times for both clumps are less than 1 Myr ($\sim 0.2$ Myr for WHL0137-zD1.1 and $\sim 0.5$ Myr for MACS0308-zD1.1).
This could indicate that the clumps are bound if their ages are much greater than 1 Myr. 
However, there is considerable uncertainty on the ages of these clumps, as stellar population ages are poorly constrained with only the rest-frame UV available to constrain our SED fits.

We can utilize a few other statistics to assess the possible fates of these clumps.
For example, theoretical models and simulations have found that clumps reaching a circular velocity of $\sim 50-100$ km s$^{-1}$ have a strong enough binding energy to be stable to supernova feedback \citep{Mandelker17,Dekel09,DekelSilk1987}.
We calculate circular velocities ($v_{circ} = \sqrt{GM/R}$) of $130 \pm 20$ for WHL0137-zD1.1, and $500 \pm 100$ for MACS0308-zD1.1. 
This suggests that these clumps are stable to disruption by supernova feedback.
Additionally, previous work has found that clumps with gas surface mass densities $\Sigma > 300 M_{\odot} \textrm{pc}^{-2}$ will be unaffected by radiation pressure feedback \citep{Mandelker17,KrumholzDekel10}.
Here, we find WHL0137-zD1.1 has a surface stellar mass density of $\Sigma_{M_*} = 4.4 \pm 0.6 \times 10^5 M_{\odot} \textrm{pc}^{-2}$, while MACS0308-zD1.1 has $\Sigma_{M_*} = 6.9 \pm 1.3 \times 10^5 M_{\odot} \textrm{pc}^{-2}$.
Given that the stellar density of a clump must be influenced by the gas density in the parent cloud, the fact that our measured stellar densities are so much greater than the threshold for stability suggests that these clumps may be stable to disruption by radiation pressure feedback.
Together, these results imply that these clumps are likely to be long-lived systems.

\begin{table*}[]
    \centering
    \begin{tabular}{c c c c c c}
        Clump & Magnification & Radius & $M_{UV}$ & SFR & $\Sigma_{\textrm{SFR}}$ \\
          & & pc & & $M_{\odot} \textrm{ yr}^{-1}$ & $M_{\odot} \textrm{ yr}^{-1} \textrm{ kpc}^{-2}$ \\
        \hline 
        Sunrise Arc & $155 \pm 13$ & & $-17.6 \pm 0.2$ & $0.54 \pm 0.06$ & \\ 
        1.1 & $215 \pm 40$ & $\leq 3.0 (3.4)$ & $-14.5 \pm 0.2$ & $0.029 \pm 0.006$ & $750 \pm 140$  \\
        1.2 & $97 \pm 9$ & $\leq 4.3 (5.5)$ & $-14.1 \pm 0.3$ & $0.021 \pm 0.005$ & $280 \pm 40$ \\
        1.3 & $63 \pm 6$ & $\leq 5.5 (8.7)$ & $-14.1 \pm 0.4$ & $0.021 \pm 0.008$ & $200 \pm 50$ \\
        1.4 & $79 \pm 8$ & $\leq 5.6 (9.9)$ & $-14.5 \pm 0.3$ & $0.03 \pm 0.009$  & $240 \pm 80$ \\
        1.5 & $160 \pm 30$ & $\leq 1.8 (2.6)$ & $-12.4 \pm 0.4$ & $0.005 \pm 0.002$ & $380 \pm 90$ \\
        1.6 & $180 \pm 50$ & $\leq 5.5 (12.4)$ & $-12.9 \pm 0.6$ & $0.007 \pm 0.004$ & $80 \pm 40$ \\
        1.7 & $250 \pm 160$ & $\leq 3.1 (4.8)$ & $-13.1 \pm 0.8$ & $0.008 \pm 0.005$ & $210 \pm 140$ \\
        Diffuse & $130 \pm 70$ & $200 \pm 60$ & $-17.3 \pm 0.6$ & $0.4 \pm 0.2$ & $2.1 \pm 1.2$ \\
        \hline 
        MACS0308-zD & & & $-20.55 \pm 0.03$ & $7.9 \pm 0.2$ &   \\
        1.1  & $22 \pm 7$ & $27 \pm 5$ & $-19.6 \pm 0.3$ & $3 \pm 1$ & $900\pm 300$  \\
        Diffuse & $19 \pm 4$ & $900\pm 200$  & $-20.4\pm 0.3$ & $7 \pm 2$ & $0.22 \pm 0.08$\\
        \hline 
        SPT0615-JD1 & $5^{+3}_{-1} $ & & $-20.73 \pm 0.03$ & $9.35 \pm 0.34$ & \\ 
        1.1 & $5^{+3}_{-1} $ & $65^{+20}_{-9}$ & $-19.0^{+0.3}_{-0.7}$ & $1.9^{+0.5}_{-1.2}$ & $100^{+60}_{-25}$ \\
        1.2 & $5^{+3}_{-1}$ & $40^{+14}_{-8}$ & $-18.3^{+0.4}_{-0.7}$ & $1.0^{+0.4}_{-0.7}$ & $130^{+80}_{-30}$  \\
        1.3 & $5^{+2}_{-1}$ & $23^{+8}_{-6}$ & $-17.7^{+0.6}_{-0.7}$ & $0.6^{+0.4}_{-0.3}$ & $230^{+90}_{-50}$  \\
        1.4 & $5^{+3}_{-1}$ & $70^{+30}_{-20}$ & $-18.6^{+0.4}_{-0.7}$ & $1.3^{+0.8}_{-0.4}$ & $60\pm 35$ \\
        1.5 & $5^{+3}_{-1}$ & $30^{+11}_{-7}$  & $-18.0^{+0.5}_{-0.8}$ & $0.8^{+0.6}_{-0.3}$ & $200^{+120}_{-50}$ \\
        Diffuse & $5^{+3}_{-1}$ & $450\pm 700$ & $-18\pm 2$ & $1 \pm 1$ & $5^{+8}_{-2}$ 

    \end{tabular}
    \caption{ Forward model results are presented for each clump, as well as the diffuse background arc where applicable. For the Sunrise Arc, radius upper limits are quoted at the 68\% (95\%) confidence interval. All other uncertainties are quoted at the $1\sigma$ level, and magnification uncertainties are propagated to derived quantities. SFR presented in this table is calculated from the delensed UV luminosity, as discussed in the text. Note for the Sunrise Arc, magnification uncertainties are calculated using only statistical variations from one lens model, and thus uncertainties are likely underestimated.} 
    \label{tab:fwd_clumps}
\end{table*}

\begin{table*}[]
    \centering
    \begin{tabular}{c c c c c c}
         Clump & SFR & $\log(M_*)$ & $\log(\textrm{sSFR})$ & Age  \\
          & $M_{\odot} \textrm{ yr}^{-1}$ & $M_{\odot}$ & yr$^-1$ &  Myr \\
          \hline
          Sunrise Arc & $0.41^{+0.14}_{-0.06}$ & $8.1^{+0.2}_{-0.3}$ & $-8.5^{+0.3}_{-0.2}$ & $400^{+300}_{-200}$ \\
          WHL0137-zD1.1 & $0.04^{+0.03}_{-0.01}$ & $7.1^{+0.3}_{-0.4}$ & $-8.5^{+0.4}_{-0.2}$ & $400^{+300}_{-250}$ \\
          \hline
          MACS0308-zD &  $6.2^{+1.8}_{-0.9}$ & $9.3^{+0.2}_{-0.4}$ & $-8.5^{+0.4}_{-0.2}$ & $400^{+270}_{-250}$ \\
          MACS0308-zD1.1 & $4.7^{+1.4}_{-0.6}$ & $9.2^{+0.2}_{-0.4}$ & $-8.5^{+0.4}_{-0.2}$ &  $400^{+300}_{-200}$ \\ 
          \hline
          SPT0615-JD1 & $4.5^{+1.7}_{-2.4}$ & $8.6\pm 0.3$ & $-7.9 \pm 0.2$ & $60^{+100}_{-40}$ \\ 
    \end{tabular}
    \caption{SED fitting results for each full arc are shown, along with individual clump fits for WHL0137-zD1.1 and MACS0308-zD1.1. All other clumps are too faint or too blended to be fit individually. Magnification uncertainties are incorporated into quoted uncertainties on all parameters.}
    \label{tab:pipes_clumps}
\end{table*}

\section{Discussion}
\label{sec:discussion}

\subsection{Compact Star Formation at High-$z$}

The star forming clumps presented in this paper are notably compact, particularly within the Sunrise Arc.
These clumps each have upper limit radii of below 12 pc, which makes them the smallest limits on clump sizes thus far observed at $z > 6$, smaller than the previous record of $r < 13$ pc presented in \cite{Vanzella19_13pc}.
These small radii put the clumps of the Sunrise Arc squarely in the regime of Young Massive Clusters (YMCs), which have been measured with radii as small as $\sim 1$ pc \citep{Portegies-zwart2010_YMCs,Ryon17}.
YMCs in turn are thought to be precursors to globular clusters \citep[e.g.,][]{Terlevich18}, however there is some debate on this topic as local YMCs have not been found to contain multiple stellar populations indicative of globular clusters \citep{BastianLardo18}.
Either way, the presence of seven highly magnified YMCs in a galaxy at $z > 6$ presents a great opportunity to study high redshift star clusters in detail with future JWST observations. 

The star clusters measured in MACS0308-zD and SPT0615-JD1 are somewhat larger than those in the Sunrise Arc, with radii measuring tens of parsecs.
This puts these clumps on the larger end of the YMC scale based solely on radius \citep{Portegies-zwart2010_YMCs}, however the stellar mass of MACS0308-zD1.1 ($\sim 10^9 M_{\odot}$ indicates that it may be comprised of multiple merging star clusters, or perhaps a dense nucleus of a merging galaxy.
It is worth noting that the clumps in SPT0615-JD1 may intrinsically be smaller than what is measured here. 
This arc is at a lower magnification ($\mu \sim 5$) than the other two, and its morphology shows fewer clearly distinguished features.
This is indicative of multiple smaller clumps blending together to form the structures we observe \citep[as in e.g.,][]{Rigby17}.
We attempt to model this blending using multiple smaller structures to model the clumps in this arc, but ultimately higher resolution imaging will be needed to better constrain the clump sizes.

We note also that, as our observations push the limits of what can be observed with \HST, our samples run up against completeness limits, as shown by dashed curves in Figure \ref{fig:sfr_radius}. 
These limits are calculated as follows. 
The vertical portion of the curves are calculated from the magnification-corrected diffraction limit, dividing the \HST WFC3/IR resolution (0.13\arcsec) by the magnification. 
For the $z=6$ curve shown, we set the magnified diffraction limit based on the brightest clump in the Sunrise Arc, magnified by a factor $\mu=215$. We note that one smaller clump exists for this sample, but its magnification is less well constrained. 
The horizontal bottom portion of the curve is calculated from the 5-$\sigma$ limiting magnitude of our observations, again corrected for the lensing magnification. 
The slanted side of the curve represents the surface brightness limit, calculated at each redshift from cosmological surface brightness dimming: $2.5\log(1+z)^4$. 
This surface brightness limit has a slope of 5 mag/dex. 
Additional discussion of these completeness limits can be found in \cite{Windhorst21_orcas}.
Our completeness limits indicate that there may be smaller, fainter structures beyond the reach of our current observations. 
In particular, we cannot rule out that we are seeing only the brightest and most highly star forming clumps, and others that better fit with the trend seen in lower-redshift clumps may exist beyond what we can currently observe.
For the clumps of SPT0615-JD, we note that all clumps appear on the edge of the completeness limit. 
This again justifies our decision to model several of the detected clumps as multiple, smaller features. 
Again in this arc, additional structure may exist beyond the current limits of our observations. 

Our forward modeling provides a measure of intrinsic luminosity, which can be used to calculate SFR via the scaling relation in \cite{MadauDickinson14}.
The resulting SFRs are rather high for each clump given their radii, indicating that these clumps host intense star formation.
To quantify this, we calculate the SFR surface density (SFRSD) using the measured radius and SFR, finding generally high SFRSDs in each clump.
In particular, clump 1.1 of MACS0308-zD has $\Sigma_{SFR} = 900 \pm 300 M_{\odot} \textrm{ yr}^{-1}\textrm{ kpc}^{-2}$, nearing the upper edge of the Kennicutt-Schmidt relation \citep{Kennicutt98b_KSlaw}, and the regime of maximal Eddington-limited star formation rate calculated by \cite{Crocker18}.
The clumps of the Sunrise Arc generally have SFRSDs of a few hundred $M_{\odot} \textrm{ yr}^{-1}\textrm{ kpc}^{-2}$, which is still quite high.
Clump 1.1 in the Sunrise Arc in particular has $\Sigma_{SFR} = 750 \pm 140$, similar to that measured for the clump D1(core) discussed in \cite{Vanzella19_13pc}.
These SFRSDs are also consistent with dense star clusters observed locally \citep[e.g.,][]{Adamo17,Ryon17}.
We note however that the $L_{UV}-$SFR relation of \cite{MadauDickinson14} may break down for the clumps of the Sunrise Arc.
These clumps show particularly low SFRs, which means that stochastic star formation effects become a greater source of uncertainty, resulting in larger scatter in the $L_{UV}-$SFR relation \citep{daSilva12_slug1,daSilva14_slug2,Vikeus2020}.

One particularly interesting question for these YMCs in the early universe is whether they remain bound, going on to form globular clusters, or if they disperse either through tidal disruption or stellar feedback.
Simulations of larger (100-1000 pc) clumps in galaxies at $z \sim 2$ have found that these structures tend to be disrupted either by stellar feedback or gravitational interactions within a few hundred Myr at most \citep[][]{Oklopcic17,MengGnedin2020}.
However, other simulations have found that more massive gravitationally bound clumps can survive long-term, forming bound globular clusters \citep{Mandelker17}.
One useful way to determine if a cluster is bound is to compare its crossing time $t_{cross}$ to its age.
The ratio of these quantities can provide a useful metric for the boundedness of star clusters, as if a cluster has survived for many crossing times, it is likely gravitationally bound. \citep{Gieles11,Bastian12}.
Our current data can only provide an estimate of the crossing time for MACS0308-zD1.1 and WHL0137-zD1.1, however both of these are less than one Myr.
Additional measurements of the stellar mass densities and circular velocities further support the interpretation of these as long-lived objects. 
Recent simulation results have also suggested that external effects within their host galaxies \citep{Rodriguez22}, though we do not directly assess these effects here.
The mass of MACS0308-zD1.1, along with the slight residual in the forward model fit, indicates that it may more likely be comprised of multiple unresolved star clusters, not just one YMC. 
However, given the available evidence, we conclude that WHL0137-zD1.1 is a YMC that appears likely to survive long-term, making it a proto-globular cluster candidate.


\section{Conclusions}
\label{sec:conclusions}

We present observations of three superlative arcs from the RELICS survey. 
These include the longest $z\sim 6$ arc (Sunrise), the brightest $z\sim 6$ arc (MACS0308-zD1), and the most distant resolved arc at $z\sim 10$ (SPT0615-JD1). 
Each of these arcs show clear substructure on scales ranging from tens of parsecs down to a few parsecs. 
Most of the clumps are likely YMCs, and they all exhibit high star formation rates compared to their small sizes.
Future observations with JWST will provide greater detail on the inner workings of these superlative lensed galaxies, including determinations of clump ages which in turn determine whether or not a clump is gravitationally bound. 
In particular, an accepted JWST Cycle 1 program (GO-2282) will obtain spectra of many star clusters within the Sunrise Arc, allowing more accurate measures of star formation rates and ages. 
Beyond that, these observations will measure metallicities, and could constrain ionization parameters for each star cluster. 
Together, these observations will better constrain the formation and environment within this $z \sim 6$ galaxy.
Similar spectroscopic observations of the other galaxies mentioned herein will be similarly beneficial, providing better constraints on all parameters presented here as well as new measurements of other physical parameters. 
These galaxies will be excellent targets for future observations with JWST as well as other observatories.

Acknowledgements:

The RELICS Hubble Treasury Program (GO 14096) and follow-up programs (GO 15842, GO 15920) 
consist of observations obtained by the NASA/ESA \emph{Hubble Space Telescope (HST)}.
Data from these {\it HST } programs were obtained from the Mikulski Archive for Space Telescopes (MAST), operated by the Space Telescope Science Institute (STScI).
Both {\it HST } and STScI are operated by the Association of Universities for Research in Astronomy, Inc.~(AURA), under NASA contract NAS 5-26555.
The {\it HST } Advanced Camera for Surveys (ACS) was developed under NASA contract NAS 5-32864.
AZ acknowledges support by Grant No. 2020750 from the United States-Israel Binational Science Foundation (BSF) and Grant No. 2109066 from the United States National Science Foundation (NSF), and by the Ministry of Science \& Technology, Israel.
EZ acknowledges funding from the Swedish National Space Agency.

\bibliography{masterbib.bib}

\end{document}